\documentclass[prb,a4paper,twocolumn,showpacs,floatfix,superscriptaddress]{
revtex4}
\usepackage{graphicx}
\usepackage{amssymb}
\usepackage{amsmath}
\usepackage{color}

\begin{document}

\title{Local attractors, degeneracy and analyticity: symmetry effects on the
locally coupled Kuramoto model}
\author{Paulo F. C. Tilles}
\affiliation{Instituto de F\'{i}sica Te\'{o}rica UNESP - Universidade Estadual
Paulista, Rua Dr. Bento Teobaldo Ferraz, 271, Bloco II, Barra Funda, 01140-070
S\~{a}o Paulo, Brazil}
\affiliation{Instituto de F\'{i}­sica de S\~ao Carlos, Universidade de S\~ao Paulo,
Caixa Postal 369, 13560-970, S\~ao Carlos SP, Brazil}
\author{Hilda A. Cerdeira}
\affiliation{Instituto de F\'{i}sica Te\'{o}rica UNESP - Universidade Estadual
Paulista, Rua Dr. Bento Teobaldo Ferraz, 271, Bloco II, Barra Funda, 01140-070
S\~{a}o Paulo, Brazil}
\author{Fernando F. Ferreira}
\affiliation{GRIFE, Escola de Arte, Ci\^encias e Humanidades,
 Universidade de S\~ao Paulo,  Av. Arlindo Bettio 1000, 03828-000 S\~ao
 Paulo, Brazil}
\pacs{05.45.Xt,05.45.Jn,05.45.-a}

%\date{\today}

\begin{abstract}

In this work we study the local coupled Kuramoto model with periodic
boundary conditions. Our main objective is to show how analytical solutions may
be obtained from symmetry assumptions, and while we proceed on our endeavor we
show apart from the existence of local attractors, some
unexpected features resulting from the symmetry properties, such as
intermittent and chaotic period phase slips, degeneracy of stable solutions and
double bifurcation composition. As a result of our
analysis, we show that stable fixed points in the synchronized region may be
obtained with just a small amount of the existent solutions, and for a class of
natural frequencies configuration we show analytical expressions for the
critical synchronization coupling as a function of the number of oscillators,
both exact and asymptotic.

\end{abstract}

\keywords{Phase coupled oscillators, Synchronization, Kuramoto model}

\pacs{05.45.Xt, 05.45.-a, 05.45.Jn}

\maketitle

\section{Introduction}

Nonlinear systems tend to surprise us with their behavior contrary to what we
expect from knowledge, intuition and reasoning, even when the problem appears to
be rather simple, as the Kuramoto model \cite{sakag} and its local version,
the locally coupled Kuramoto model (LCKM).
The applicability of the model for a variety of systems that range from physics
\cite{wcs,daniels}, data mining \cite{miyano}, neurosciences \cite{cumin,frank}, robotics \cite{moioli}, animal gaits \cite{collins,rogaey}, antennas \cite{rogaey}
and others make it important to understand the nature of unexpected results.
Although the connection between the LCKM and applications is reasonably well
established, the theoretical understanding is restricted to a few works
\cite{daniels,liu,tilles,bridges,hassan01,zheng,anand,rogge,hassan02,ochab}.
Recently we studied the LCKM above the
synchronization transition with periodic boundary conditions \cite{tilles}.
On the other hand we found a richness
in the
solutions which was not expected, mostly because intuition tells us that a ring
of oscillators should behave as a chain for large systems. We studied how they
are born and the region of phase space where the solutions exist. The complexity
of the problem makes it extremely hard to analyze therefore we shall start
looking at small systems with increasing difficulty to see if we can obtain some
results from which we can infer large $N$ behavior, particularly when symmetry
in the frequency distributions is assumed\cite{golubitsky}.

Our objective is to calculate the behavior of the solutions for several cases
where the distribution of frequencies have well determined symmetries. We will
study how
these solutions appear and follow them as the system size increases. To do that
we shall start with the smallest system (N=3) which has already received the
attention of Maistrenko et al \cite{maistrenko,maistrenko2} and Ashwin et al
\cite{ashwin}, and we are going to
increase the size in the search of general properties. Obviously we shall not
attempt to have an encyclopedic coverage but will show the difficulties
encountered in solving the general problem with any arbitrary distribution. In
doing so we have come across some results on phase slip which happens at the
transition to full synchronization which seem to contradict previous results for
specific symmetries \cite{hassan01,zheng}, which we shall discuss along the
text.

The manuscript follows a route to show the loss of
 analyticity when one goes from small to large systems for some specific
symmetry classes. To acomplish that it is organized as follows: the model is
introduced in Section II;
in Section III we study small systems. We start in Section III
A with a well known case,
that of three oscillators, where
we obtain for the first time the analytic result for the dependence of the
solutions on the coupling constant above
synchronization; after that, in Section III B, we consider a symmetric $N=4$
case and show a general method for solving the equations for any number of
oscillators, along with its difficulties. In Section IV we
increase the system size and impose specular
symmetry, with two natural frequencies. Under these constraints we  discuss some
properties for this class using a system of $N=6$ oscillators, obtaining
analytical results.
The analyticity
limits of the model are explored in section V when the frequencies
are obtained from a random distribution while keeping the specular
symmetry. In this case we present analytical
solutions only for $K_s$.
In section VI we show some examples on how organization on the
natural frequencies may lead to analytical (asymptotic) expressions for the
critical synchronization coupling. A summary of the results and some possible
further extensions are left for the final section.

Before ending the introduction we define some of the notation used
throughout the text. When treating the bifurcations of the model we are not
interested in classifying them, but on the characterization of the fixed
point's stability.
 With that in mind, although it may seem as if we try to
redefine well known bifurcations we name them according to the type of fixed
points created at them to facilitate identification;
for the fixed point classification, we call a saddle every solution containing
both positive and negative eigenvalues of the Jacobian matrix and stable
(unstable) node when every eigenvalue is negative (positive), regardless of
their
imaginary parts.
\section{Periodic Boundary Conditions: effects of symmetry}

The LCKM under periodic boundary conditions presents a rich
landscape of solutions as discussed in reference \cite{tilles}. From now on we
shall follow the nomenclature used in that reference. The system is described by
the set of equations:
\begin{equation}
\dot{\theta}_{n}=\omega _{n}+K\left[ \sin \left( \theta _{n-1}-\theta_{n}\right)
+\sin \left( \theta _{n+1}-\theta _{n}\right) \right], \label{5.015}
\end{equation}
for $n=1,...,N$, where $\omega _{n}\in \left\{ \omega \right\}_{N}$ is the set
of natural frequencies. The topology of the ring is defined by the conditions
$\theta _{N+1}=\theta_{1}$ and $\theta_{0}=\theta _{N}$. Alternatively, the
system (\ref{5.015}) may also be written in terms of phase differences $\phi_{n}
= \theta _{n}-\theta _{n+1}$,
\begin{equation}
\dot{\phi}_{n}=\omega _{n} -\omega _{n+1}+K\left[ \sin{\phi_{n-1}
-2\sin{\phi_{n}} +\sin{\phi_{n+1}}}
\right], \label{ad01}
\end{equation}
subject to the natural identity $\sum_{n=1}^{N} \phi_{n}=0$, with the periodic
boundary conditions satisfied by $\phi_{0}=\phi_{N}$ and $\phi_{N+1}=\phi_{1}$.
Similarly to the case
of the chain \cite{strog01} there is a minimum  value of the coupling constant
$K$ for which
the system synchronizes to a common frequency $\Omega =
\frac{1}{N} \sum_{j=1}^{N}\omega _{j}$. Under full synchronization the set of
equations (\ref{5.015}) can be written as
\begin{equation}
\frac{\Omega -\omega _{n}}{K}=\sin \phi _{n-1}-\sin \phi _{n}, \qquad n=1,...,N,
 \label{5.016}
\end{equation}
where $\phi _{n}=\theta _{n}-\theta _{n+1}$. The condition for the phase
differences $\phi _{n}\left( K,\left\{ \omega \right\} \right)$ to lock depends
on the number of oscillators $N$ and the value of the coupling constant in the
region $K\geq K_{s}$ ($K_s$ is a unique fixed point that
represents the onset of synchronization). In the synchronized state every
variable
$\phi _{n}$ can be written in terms of one $\phi _{n^{\ast }}$
arbitrarily chosen:
\begin{subequations}
\label{5.017}
\begin{equation}
\sin{\phi_{n}} = \sin{\phi_{n^{\ast}}} + \frac{1}{K}
\sum_{j=n+1}^{n^{\ast}}\left( \Omega -\omega _{j}\right), \label{5.017a}
\end{equation}
for $n=1,...,n^{\ast }-1$ and
\begin{equation}
\sin{\phi_{n}} = \sin{\phi_{n^{\ast}}} -\frac{1}{K}\sum_{j=n^{\ast}+1}^{n}\left(
\Omega -\omega _{j}\right), \label{5.017b}
\end{equation}
\end{subequations}
for $n=n^{\ast }+1,...,N-1$, as long as we keep the right hand side on both
equations in the interval [-1,1]. Since the identity $\sum_{j=1}^{N}\phi _{j}=0$
allows us to write $\phi _{N}$ as a sum of all the other phases $\phi _{n}$,
with $n=1,...,N-1$, the set of equations (\ref{5.016}) is reduced to a single
equation on two variables $\left( \phi _{n^{\ast}},K\right)$:
\begin{equation}
\sin \left( \phi _{n^{\ast }}+\sum_{n\neq n^{\ast }}^{N-1}\phi _{n}\right) +\sin
\phi _{n^{\ast }}=\frac{\sum_{j=1}^{n^{\ast }}\left(\omega_{j} - \Omega
\right)}{K},  \label{5.018}
\end{equation}
where $\phi_{n}$ are determined by equations \ref{5.017a} and
\ref{5.017b} .

Ochab and Gora \cite{ochab}  found an equation similar to
equation (\ref{5.018}) for the topology of a ring. They introduced a parameter
$p=-\sin(\phi_{N-1})$ and vary it in the interval $[-1,1]$ to find numerical
solutions for different winding numbers.
They found only one branch of solutions for each winding number per $K$. In this
work we treat only
winding number $m=0$, which we showed in reference \cite{tilles} that it has
multiple solutions above synchronization.  The description of the solutions
equation (\ref{5.018}) as a function of $K$ is not a simple extension
of that of the chain since for the ring topology the equations for the phase
differences are not independent.
It is necessary to analyze simple particular
cases where the small size, or particular symmetries, will give us a hint on how
to proceed to the general case of randomly selected natural frequencies. This is
what we shall do in the following sections.

\section{Small systems analysis}

In this section we make an analysis of small systems,
particularly the $N=3$ and $N=4$. We show two approaches for obtaining
analytical solutions, and discuss some of the
problems within this description when extended to higher number of oscillators.
Unless specified on the text, from now on the values assumed by
$\arcsin \left( x\right)$ are to be considered only on the interval $\left[
-\pi/2, \pi/2
\right]$.

\subsection{First non trivial case: N=3}

The first non trivial case is a system of three oscillators and natural
frequencies $\omega_{1} = -\omega_{3}=\omega$ and $\omega_{2}=0$. Clearly
$\Omega = 0$ and the synchronized region is described by:
\begin{subequations}
\label{5.019}
\begin{eqnarray}
\sin{\left( \phi_{1} + \phi_{2}\right)} + \sin{\phi_{1}} &=& \frac{\omega}{K},
\label{5.019a} \\
\sin{\phi_{2}} - \sin{\phi_{1}} &=& 0. \label{5.019b}
\end{eqnarray}
\end{subequations}
Since (\ref{5.019b}) has only two solutions,
\begin{equation}
\phi^{I}_{2} = \phi_{1}, \qquad \phi^{II}_{2} = \pi - \phi_{1},  \label{5.020}
\end{equation}
it is possible to use only one phase difference $\phi_{1} = \phi$ and rewrite
(\ref{5.019}) as a single equation $K\left(\phi \right)$ for each solution:
\begin{equation}
K^{I} \left( \phi \right)=\frac{\omega}{\sin \phi + \sin 2\phi}, \qquad K^{II}
\left( \phi \right)= \frac{\omega}{\sin{\phi}}. \label{5.021}
\end{equation}

The function $K^{II} \left( \phi \right)$ has the same structure
as a chain of oscillators, with just a
single minimum at $K^{II} = \omega$ with $\sin{\phi=1}$. On the other hand
$K^{I} \left( \phi \right)$ has two minima for $K\geq0$. If we define $z=\sin
\phi$ these minima can be written as $K^{I}_{\pm}=f\left(z_{\pm}^{*} \right)$,
i.e.:
\begin{equation}
K^{I}_{\pm}=\frac{\omega}{z_{\pm}^{*} \left(1 \pm 2 \sqrt{1-z_{\pm}^{*2}}
\right)}, \label{5.022}
\end{equation}
where
\begin{equation}
z_{+}^{*}=\sqrt \frac{15+\sqrt{33}}{32}, \qquad z_{-}^{*}=-\sqrt
\frac{15-\sqrt{33}}{32}. \label{5.023}
\end{equation}

Each of the three minima gives birth to a pair of solutions for $K \geq K_{s}$,
and their stability is characterized by the eigenvalues of the Jacobian matrix.
When we write the equations of motion as a function of the
phase differences
\begin{subequations}
\label{5.024}
\begin{eqnarray}
\hspace{-0.6cm} \dot{\phi}_{1} &=& \omega - K \left[2 \sin{\phi_{1}} -
\sin{\phi_{2}} + \sin{\left(\phi_{1} + \phi_{2} \right)}\right], \label{5.024a}
\\
\hspace{-0.6cm} \dot{\phi}_{2} &=& \omega + K \left[\sin{\phi_{1}} - 2
\sin{\phi_{2}} - \sin{\left(\phi_{1} + \phi_{2} \right)}\right], \label{5.024b}
\end{eqnarray}
\end{subequations}
the eigenvalues for each type of solution are defined by
\begin{eqnarray}
\hspace{-0.4cm} \lambda^{I}_{\pm} \left(\phi \right)&=& K \left[-2\cos\phi
-\cos{2\phi} \pm \left(
\cos\phi -\cos{2\phi}\right) \right], \label{5.026} \\
\hspace{-0.4cm} \lambda^{II}_{\pm} \left(\phi \right)&=&  K \left(1
\pm \sqrt{1 +3\cos^{2}{\phi}}  \right). \label{5.027}
\end{eqnarray}
\begin{figure}[!t]

\centering
\includegraphics[width=1.0\linewidth]{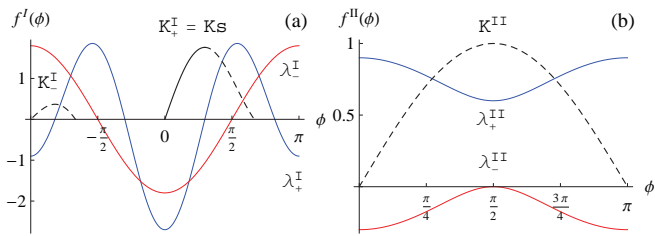}
\caption{(Color online) (a) Plot of the functions $f^{I}=1/K^{I}_{\pm}
\left(\phi
\right)$ where the stable solutions are shown in solid black lines and the
unstable solutions in dashed lines and $\lambda^{I}_{\pm}\left(\phi \right)$
(thin blue and red lines) for $\omega = 1$ and the symmetric N=3 system defined by equations (6).
Saddle-(stable)node bifurcation at
$K^{I}_{+} = K_s$, the solutions are stable on the left of the bifurcation and
one of the solutions is unstable, on the right. At the minimum of $K^{I}_{-}$
there is a saddle-(unstable)node bifurcation. (b) For $f^{II}=1/K^{II}
\left(\phi
\right)$ and $\lambda^{II}_{\pm}\left(\phi \right)$ we see a (saddle-saddle
bifurcation at the minimum of $K^{II} \left(\phi \right)$).}
\label{Cap05_fig02}
\end{figure}
To analyze the stability of the system, we plot the functions
$f^{I}=1/K^{I}_{\pm}
\left(\phi \right)$ and $f^{II}=1/K^{II} \left(\phi \right)$, with the
corresponding
eigenvalues of the Jacobian matrix versus $\phi$ in figure \ref{Cap05_fig02}. We
notice that the first maximum at $K^{I}_{+}$ represents a saddle-(stable)node
bifurcation:  two solutions are born, one stable with $\phi \rightarrow 0$ and
one unstable with $\phi \rightarrow 2 \pi/3$ in the limit $K\rightarrow \infty$.
The solutions generated at  the minima $K^{I}_{-}$  (figure \ref{Cap05_fig02}a)
and $K^{II}$ (figure \ref{Cap05_fig02}b) are unstable, since the real part of
$\lambda^{I}_{-}$ and $\lambda^{II}_{+}$ is always positive: $K^{I}_{-}$ is a
saddle-(unstable)node bifurcation and $K^{II}$ is a saddle-saddle bifurcation.

For this system with just a few oscillators it is not necessary to impose
symmetry properties on the natural frequencies in order to obtain an analytical
description of the synchronized region. If we consider the natural frequencies
to be randomly chosen, it is always possible to assign positive values to
$\omega_{1}$ and $\omega_{2}$ and set $\omega_{3} =-\omega_{1}-\omega_{2}$ (for
$\Omega=0$). With this configuration the fixed point solutions are described by
the equations:
\begin{subequations}
\label{5.031}
\begin{eqnarray}
\sin{\left( \phi_{1} + \phi_{2}\right)} + \sin{\phi_{2}} &=&
\frac{\omega_{1}+\omega_{2}}{K}, \label{5.031a} \\
\sin{\phi_{1}} - \sin{\phi_{2}} &=& - \frac{\omega_{2}}{K}. \label{5.031b}
\end{eqnarray}
\end{subequations}
A suitable manipulation of equations (\ref{5.031}) enable us to write a single
equation for $\phi_{2}$,
\begin{eqnarray}
&& \hspace{-0.8cm} \sin^{2}{\phi_{2}}\left[1-\left(\sin{\phi_{2}}-\omega_{2}x
\right)^{2} \right] = \nonumber \\
&& \hspace{-0.5cm} \left[\left(\omega_{1}+\omega_{2} \right)x -\sin{\phi_{2}}
-\left(\sin{\phi_{2}}-\omega_{2}x\right)\cos{\phi_{2}}\right]^{2}, \label{5.034}
\end{eqnarray}
where $x=K^{-1}$. Since (\ref{5.034}) is a polynomial of second order in $x$, we
find solutions in the same fashion as equation (\ref{5.021}),
i.e., functional forms $K^{-1}\left(\phi_{2}\right)$ describing all the fixed
points on the synchronized region:
\begin{subequations}
\label{5.035}
\begin{eqnarray}
\hspace{-0.5cm} K_{\pm}^{-1}\left(\phi_{2}\right) &=& \nonumber \\
&& \hspace{-1.0cm} \frac{\left(\omega_{1} +2\omega_{2} \right) \sin{\phi_{2}}
\left(1 +\cos{\phi_{2}}\right) \pm \sqrt{k\left(\phi_{2}\right)}
}{\omega_{1}^{2} +2\omega_{2} \left(\omega_{1} +\omega_{2}\right) \left(1
+\cos{\phi_{2}}\right)}, \label{5.035a} \\
\hspace{-0.5cm} k\left(\phi_{2}\right) &=& \sin^{2}{\phi_{2}}
\left[\omega_{1}^{2} \cos^{2}{\phi_{2}} \right. \nonumber \\
&& \hspace{-0.5cm} \left.+2\omega_{2}\left(\omega_{1}+\omega_{2}\right)
\left(1+\cos{\phi_{2}}\right) \right]. \label{5.035b}
\end{eqnarray}
\end{subequations}
For the symmetric case the fixed points are those shown in
figure \ref{Cap05_fig02}.

\begin{figure}[!t]
\centering

\includegraphics[width=1.0\linewidth]{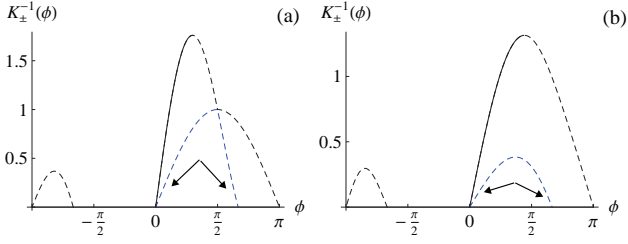}
\caption{(Color online) General solution $K_{\pm}^{-1} \left( \phi_{2}\right)$
representing the fixed points of the $N=3$ system. Stable (unstable) branches
are represented by solid (dashed) lines. (a) Complete solution for the symmetric
case shown in figure \ref{Cap05_fig02}, where it is possible to see a pitchfork
bifurcation at $\phi_{2}=\pi/2$. (b) General structure of the synchronized
region for a configuration of natural frequencies without symmetries:
$\omega_{1} =1$ and $\omega_{2} =1/2$. The six fixed points solutions come from
three different bifurcations.
The branches that correspond to $K_{-}^{-1} $ are indicated by arrows. The
others branches correspond to $K_{+}^{-1} $  }
\label{Fig02}
\end{figure}

It is now possible to see that the $K^{II}$ solutions
are born from the rightmost branch of $K^{I}$ (figure \ref{Fig02}a), which lead
to the pitchfork bifurcation description by Maistrenko \emph{et al.}
\cite{maistrenko}.  A general configuration of natural frequencies generates the
structure of fixed points shown in figure \ref{Fig02}b, where the maxima of
$K_{\pm}^{-1}\left(\phi_{2}\right)$ appear separate, thus characterizing the
presence of three bifurcations on the phase space. Unfortunately, the analytic
expression for $K_{s} \left(\omega_{1},\omega_{2} \right)$ does not admit a
simple representation and to best of our knowledge it has not yet been
reported.

\subsection{N=4 with specular symmetry}

Based on the same assumptions as the previous case, we shall start analyzing the
ring with
$N=4$ oscillators where the configuration of frequencies follow a
prescribed symmetry. It is worth mentioning that
Maistrenko et al. \cite{maistrenko2} have already studied the case of N=4 for
the full connected Kuramoto model. We treat here the case when the oscillators
present specular
symmetry on the natural frequencies, with $\omega_{3}=-\omega_{2}$,
$\omega_{4}=-\omega_{1}$, and both $\omega_{1}$ and $\omega_{2}$ positives.

The symmetry in the equations imposes
$\sin{\phi_{1}}=\sin{\phi_{3}}$, with solutions
\begin{equation}
\phi_{3}^{I} = \phi_{1}, \qquad \phi_{3}^{II} = \pi - \phi_{1}. \label{5.036}
\end{equation}
For the $\phi_{3}^{II}$ case, the equations in the
synchronized state,
\begin{equation}
\sin{\left(\pi + \phi_{2} \right)} +\sin{\phi_{1}} = \frac{\omega_{1}}{K}, \quad
\sin{\phi_{2}} - \sin{\phi_{1}} = \frac{\omega_{2}}{K}, \label{5.037}
\end{equation}
permit the existence of a solution only if $\omega_{2}=-\omega_{1}$. As this
case belongs to a more general scenario described in the next section, we will
consider only the $\phi_{3}^{I}$ solution, whose synchronized region is
described by the equations:
\begin{subequations}
\label{5.038}
\begin{eqnarray}
\sin{\left(2 \phi_{1} + \phi_{2} \right)} +\sin{\phi_{2}} &=& \frac{\omega_{1} +
\omega_{2} }{K}, \label{5.038a} \\
\sin{\phi_{2}} - \sin{\phi_{1}} &=& \frac{\omega_{2}}{K}. \label{5.038b}
\end{eqnarray}
\end{subequations}
After some manipulation, equation (\ref{5.038a}) can be written as
\begin{eqnarray}
&& \hspace{-1.5cm} \sin^{2}{\phi_{1}} \left(1 - \sin^{2}{\phi_{1}} \right)
\left(1 -
\sin^{2}{\phi_{2}} \right) = \nonumber \\
&& \left[ \frac{\omega_{1} + \omega_{2} }{2K}
-\sin{\phi_{2}} \left(1 - \sin^{2}{\phi_{1}} \right) \right]^{2}. \label{5.039}
\end{eqnarray}
If we would like to search for a solution of the form $K^{-1}
\left(\phi_{2}\right)$, as it was made in the case $N=3$, equation (\ref{5.039})
shows that it is necessary to obtain the roots of a fourth order polynomial.
However, a more accurate investigation shows that if we define $z =
\sin{\phi_{2}}$, it is possible to write a third order polynomial,
\begin{eqnarray}
&& \hspace{-0.5cm} z^{3} \left(y_{2}-y_{1} \right) - z^{2} y_{2} \left(3y_{2} -
2y_{1}\right)
+ z \left(3y_{2}^{3} -y_{1}y_{2}^{2} +y_{1}-y_{2} \right) \nonumber \\
&& \hspace{1cm} - \frac{1}{4} \left(4y_{2}^{4} -3y_{2}^{2} +y_{1}^{2}
+2y_{1}y_{2}  \right) = 0, \label{5.040}
\end{eqnarray}
where $y_{1,2}=\omega_{1,2}/K$. On this slightly different approach, the fixed
point solutions come as the roots of equation (\ref{5.040}), namely,
\begin{widetext}
\begin{subequations}
\label{5.041}
\begin{eqnarray}
r_{1} \left( K\right) &=& \frac{2 \omega_{1} \omega_{2} - 3\omega_{2}^{2}}{3 K
\left(\omega_{1} - \omega_{2} \right)} + \frac{4 \omega_{1}^{2} \omega_{2}^{2} +
12K^{2} \left(\omega_{1}-\omega_{2}
 \right)^{2} + K^{4}
f_{0}^{2/3}\left(K\right)}{6K^{3}\left(\omega_{1}-\omega_{2} \right)
f_{0}^{1/3}\left(K\right) }, \label{5.041a} \\
r_{2}^{\pm} \left( K\right) &=& \frac{2 \omega_{1} \omega_{2} -
3\omega_{2}^{2}}{3 K \left(\omega_{1} - \omega_{2} \right)} + \frac{-4\left(1
\pm i\sqrt{3}\right)\left[3K^{2}\left(\omega_{1}-
\omega_{2} \right)^{2} +\omega_{1}^{2} \omega_{2}^{2}\right] + \left(-1 \pm
i\sqrt{3}\right) K^{4} f_{0}^{2/3}
\left(K\right)}{12K^{3}\left(\omega_{1}-\omega_{2} \right) f_{0}^{1/3}
\left(K\right)}, \label{5.041b}
\end{eqnarray}
\end{subequations}
where the function $f_{0}\left(K\right)$ is defined by
\begin{subequations}
\label{5.042}
\begin{eqnarray}
f_{0}\left(K \right) &=& f_{1}\left(K \right) + 3\sqrt{3} \sqrt{f_{2}\left(K
\right)}, \label{5.042a} \\
f_{1}\left(K \right) &=& - 8 \frac{\omega_{1}^{3}\omega_{2}^{3}}{K^{6}} - 9
\frac{\left(\omega_{1}-\omega_{2} \right)^{2} \left(3\omega_{1}^{2}
-2\omega_{1}\omega_{2}+3\omega_{2}^{2}
 \right)}{K^{4}} \label{5.042b} \\
f_{2}\left(K \right) &=& 16 \frac{\omega_{1}^{3} \omega_{2}^{3}
\left(\omega_{1}-\omega_{2} \right)^{4}}{K^{10}} +
\frac{\left(\omega_{1}-\omega_{2}\right)^{4} \left(27\omega_{1}^{4}
-36\omega_{1}^{3}\omega_{2}+2\omega_{1}^{2}\omega_{2}^{2}-36\omega_{1}\omega_{2}
^{3} +27\omega_{2}^{4} \right)}{K^{8}} -64 \frac{
\left(\omega_{1}-\omega_{2}\right)^{6}}{K^{6}}. \label{5.042c}
\end{eqnarray}
\end{subequations}
\end{widetext}

The roots $r_{1} \left(K\right)$ and $r_{2}^{-}\left(K\right)$ give the locked
solutions of $\sin{\phi_{2}}\left(K\right)$ for $K \geq K_{s}$, but as $\left|
r_{2}^{+} \right|$ is always greater than $1$ this root is not a valid
representation of $\sin{\phi_{2}}$. From these solutions it is possible to build
a bifurcation diagram for the system once we know the stability of the
solutions. Bearing in mind that we are searching for simple analytical
expressions, we will consider a reduction of the phase space dynamics. With a
proper set of initial conditions, i.e., $\phi_{3} \left(t=0 \right) = \phi_{1}
\left(t=0 \right)$, with $\phi_{1} \left(t=0 \right)$ and $\phi_{2} \left(t=0
\right)$ randomly generated, the dynamics is set to occur in a 2-dimensional
symmetric manifold. Now, if we follow the same procedure of the previous section
and write the equations of motion for the phase differences $\phi_{n}$,
\begin{subequations}
\label{5.043}
\begin{eqnarray}
\hspace{-0.3cm} \dot{\phi}_{1} &=& \omega_{1} - \omega_{2} \nonumber \\
&& - K\left[2 \sin{\phi_{1}} -\sin{\phi_{2}} + \sin{\left(2\phi_{1}+\phi_{2}
\right)} \right], \label{5.043a} \\
\hspace{-0.3cm} \dot{\phi}_{2} &=& 2\omega_{2} +2\left(\sin{\phi_{1}}
-\sin{\phi_{2}} \right), \label{5.043b}
\end{eqnarray}
\end{subequations}
we obtain a closed analytical expression for the eigenvalues of
the Jacobian matrix,
\begin{equation}
\lambda_{\pm} = \frac{\tau \pm \sqrt{\tau^{2}-4\Delta}}{2}, \label{5.044}
\end{equation}
where the trace $\tau$ and the determinant $\Delta$ of the matrix are given by
\begin{subequations}
\label{5.045}
\begin{eqnarray}
\tau &=& -2K \cos{\phi_{1}}\left[1 + 2\cos{\left(\phi_{1}+\phi_{2} \right)}
\right], \label{5.045a} \\
\Delta &=& -4K^{2} \cos^{2}{\left(\frac{\phi_{1}+\phi_{2}}{2} \right)} \left[1
-\cos{2\phi_{1}} \right. \nonumber \\
&& \left. -2\cos{\left(\phi_{1}+\phi_{2} \right)}\right]. \label{5.045b}
\end{eqnarray}
\end{subequations}
\begin{figure}[!t]
\centering
\includegraphics[width=1.0\linewidth]{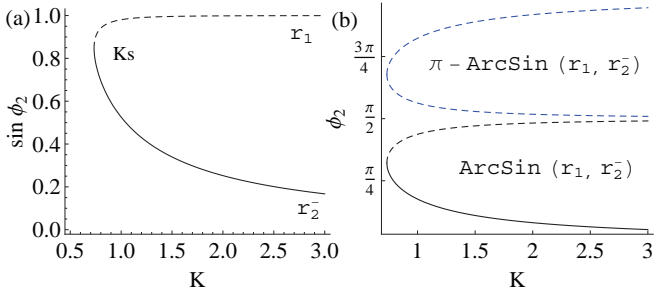}
\caption{(Color online) Roots $r_{1} \left(K\right)$ and
$r_{2}^{-}\left(K\right)$ of (\ref{5.040}) representing the bifurcation diagram
in the synchronized region for $\omega_{1}=1$ and $\omega_{2}=1/3$ for the system with N=4 defined by equations (19). Full
(dashed) lines represent stable (unstable) solutions. a) Bifurcation diagram
projected on the space $\sin{\phi_{2}} \times K$ showing the bifurcation at
$K_{s}$ where two solutions are born: one stable with $\sin{\phi_{2}}\rightarrow
0$ and another unstable with $\sin{\phi_{2}}\rightarrow 1$ in the limit $K
\rightarrow \infty$. b) Bifurcation diagram showing the simultaneous birth of
the solutions in different regions of phase space at $K_{s}$, a
saddle-(stable)node in the region $\cos{\phi_{2}}>0$ and an
saddle-(unstable)node for $\cos{\phi_{2}}<0$.}\label{Cap05_fig05}
\end{figure}
\begin{figure}[!t]
\centering
\includegraphics[width=1.0\linewidth]{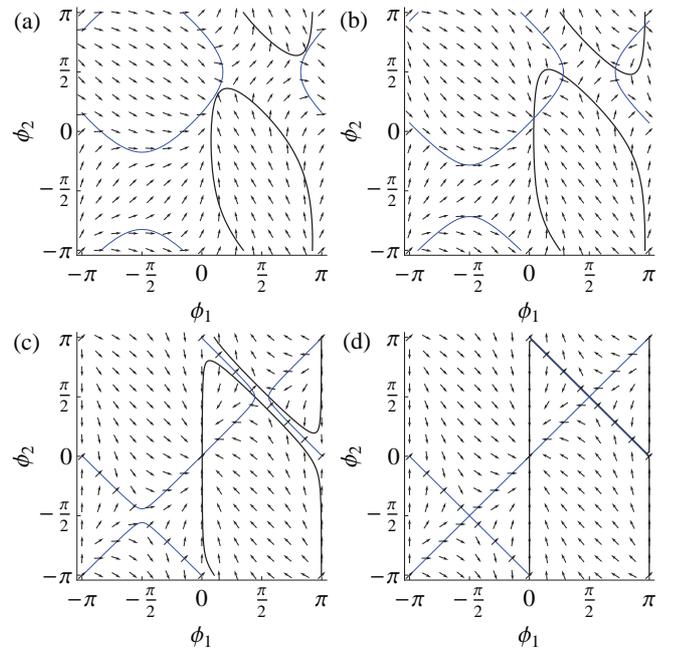}
\caption{(Color online) Phase space showing the evolution of the fixed points as
a function of $K$ for a fixed configuration of natural frequencies $\omega_{1}=1$ and $\omega_{2}=1/3$ in the system with $N=4$. Black(thick)
 and blue(thin) lines correspond to
$\dot{\phi}_{1} = 0$ ($\dot{\phi}_{2} = 0$).
a) $K=0.7$: just before $K_{s}$, we can see the simultaneous bifurcations. b)
$K=1.5$: stable node and saddle created in the region with
$\cos{\phi_{2}}>0$ (central part of the figure), and saddle and unstable node
created in the region $\cos{\phi_{2}}<0$, in the upper part of the figure.
c) $K=20.0$:  for large values of $K$ the stable node moves towards
$\left(\phi_{1},\phi_{2}\right)=\left(0,0\right)$, the two unstable saddles go
towards $\left(\pi/2,\pi/2\right)$ to merge into a unique saddle.
d) In the limit $K \rightarrow \infty$, the unstable node moves towards
$\left(\pi,\pi\right)$ while the curves deform to be a saddle at
 $\left(0,\pi\right)$, when $K\rightarrow \infty$.} \label{Cap05_fig06}
\end{figure}

At this point we find a problem: the locked solutions are given in terms of
$\sin{\phi_{2}}$ while the eigenvalues of the Jacobian depend on $\phi_{1}$ and
$\phi_{2}$. A way out of this conflict can be reached if one
notices that for each root in (\ref{5.041}) there are two solutions for
$\phi_{2}$:
\begin{subequations}
\label{5.046}
\begin{eqnarray}
&& \hspace{-1.2cm} \phi_{2}^{I} \left(r_{1} \right) = \arcsin{r_{1}}, \quad
\phi_{2}^{II} \left(r_{1} \right) = \pi - \arcsin{r_{1}},
\label{5.046a} \\
&& \hspace{-1.2cm} \phi_{2}^{I} \left(r_{2}^{-} \right) = \arcsin{r_{2}^{-}},
\quad
\phi_{2}^{II} \left(r_{2}^{-} \right) = \pi - \arcsin{r_{2}^{-}}.
\label{5.046b}
\end{eqnarray}
\end{subequations}
Since $r_{1}$ and $r_{2}^{-}$ do not depend on $\phi_{1}$, it is necessary to go
back to equations (\ref{5.038}) to determine the dependency of
$\phi_{1}$ on $\phi_{2}$ for each solution of (\ref{5.046}). A numerical
analysis of the equations shows that $\phi_{1} =
\arcsin{\left(\sin{\phi_{2}}-\frac{\omega_{2}}{K} \right)}$ leads to the
solutions $\phi_{2}^{I} \left(r_{1} \right)$ and $\phi_{2}^{I} \left(r_{2}^{-}
\right)$, while $\phi_{1} = \pi -
\arcsin{\left(\sin{\phi_{2}}-\frac{\omega_{2}}{K} \right)}$ corresponds to the
complementary branches  $\phi_{2}^{II} \left(r_{1} \right)$ and $\phi_{2}^{II}
\left(r_{2}^{-} \right)$.

As a result of this analysis it is possible to observe that, while the graph
$\sin{\phi_{2}}\times K$ shows $K_{s}$ as a bifurcation giving birth to two
solutions (figure \ref{Cap05_fig05}a), the identity $\sin{\phi} = \sin{\left(\pi
- \phi \right)}$ implies that for each branch of $\sin{\phi_{2}}$ there are
actually two fixed points in the $\phi_{2} \times K$ space (figure
\ref{Cap05_fig05}b). The unexpected (though straightforward) conclusion is the
feature that the $K_{s}$ bifurcation occurs
simultaneously in two regions of the phase space, but the fixed points do not
share the stability properties: in the region defined by
$\cos{\phi_{2}}>0$ it bifurcates into one stable $\phi_{2}^{I}
\left(r_{1}\right)$ and one unstable branch $\phi_{2}^{I}
\left(r_{2}^{-}\right)$, meanwhile for $\cos{\phi_{2}}<0$ it has
two unstable solutions, a saddle $\phi_{2}^{II} \left(r_{2}^{-}\right)$ and an
unstable node $\phi_{2}^{II} \left(r_{1}\right)$.

Now it becomes evident why we chose a reduced subspace for the dynamics: due to
its lower dimensionality it is possible to visualize the evolution of the
nullclines and the creation of the fixed points in the phase space, as shown in
figure \ref{Cap05_fig06}. Just before the bifurcation at $K_{s}$ it is possible
to visualize the two simultaneous fixed points being formed at
$\left(\phi_{1},\phi_{2} \right) \approx \left(0.4,1\right)$ and
$\left(\phi_{1},\phi_{2} \right) \approx \left(2.5,2.5\right)$ (figure
\ref{Cap05_fig06}a). Figure \ref{Cap05_fig06}b ($K > K_{s}$) shows the stable
node at $\left(\phi_{1},\phi_{2} \right) \approx \left(0,0.5\right)$ and the
saddle at $\left(\phi_{1},\phi_{2} \right) \approx \left(1,1.4\right)$ born in
the region $\cos{\phi_{2}}>0$, as well as the saddle at $\left(\phi_{1},\phi_{2}
\right) \approx \left(2.5,1.5\right)$ and the unstable node at
$\left(\phi_{1},\phi_{2} \right) \approx \left(3,2.5\right)$. In the limit $K
\rightarrow \infty$ the stable node shifts towards $\left(0,0\right)$, the two
saddles move towards $\left(\pi/2,\pi/2\right)$ until they merge into a unique
saddle, and the unstable node shifts in the direction of $\left(\pi,\pi\right)$
(figures \ref{Cap05_fig06}c and \ref{Cap05_fig06}d).

For this specific system the reduced manifold contains all the solutions of the
complete 3-dimensional phase space, and the stability of the fixed points does
not suffer any alteration when one reduces the dimension of the dynamics.
However this is not the general behavior for the LCKM when symmetries on the
natural frequencies are taking into account, as we will show in the next
section.

The value of the coupling constant at the synchronization transition can be
obtained as the minimum of the function $K\left(\sin{\phi_{2}} \right) =
K\left(z \right)$, such that $\partial_{z} K\left( z \right)
\vert_{z=z^{\ast}} = 0$ implicitly calculated in (\ref{5.040}) gives two
solutions
\begin{equation}
z^{\ast}_{\pm} = \frac{\omega_{2} \left(2\omega_{1}-3\omega_{2} \right) \pm
\sqrt{\omega_{1}^{2}\omega_{2}^{2} + 3
K^{2}\left(\omega_{1}-\omega_{2}\right)^{2}}}{3 K
\left(\omega_{1} -\omega_{2} \right)}. \label{5.047}
\end{equation}
Once we take into account the two roots $z^{\ast}_{\pm}$ in (\ref{5.040}) we
obtain four
solutions:
\begin{equation}
K_{s}^{\pm,\pm} \left(\omega_{1},\omega_{2} \right) = \pm
\frac{1}{8\sqrt{2}\left|\omega_{1}-\omega_{2} \right|} \sqrt{k_{0}^{\pm}
\left(\omega_{1},\omega_{2} \right)}, \label{5.048}
\end{equation}
where the function $k_{0}^{\pm} \left( \omega_{1},\omega_{2} \right)$ is defined
as
\begin{widetext}
\begin{subequations}
\label{5.049}
\begin{eqnarray}
k_{0}^{\pm} \left( \omega_{1},\omega_{2} \right)  &=& 27\omega_{1}^{4}
-9\omega_{1}^{3} \left(4 \omega_{2} \pm \sqrt{k_{1} \left( \omega_{1},\omega_{2}
\right)} \right) +
\omega_{2}\omega_{1}^{2} \left(2\omega_{2} \pm 5 \sqrt{k_{1} \left(
\omega_{1},\omega_{2} \right)} \right) -
\omega_{1}\omega_{2}^{2}\left(36\omega_{2} \mp 5\sqrt{k_{1}
 \left( \omega_{1},\omega_{2} \right)} \right) \nonumber \\
&& + 9\omega_{2}^{3} \left(3\omega_{2} \mp
\sqrt{k_{1} \left( \omega_{1},\omega_{2} \right)}\right), \label{5.049a} \\
k_{1} \left( \omega_{1},\omega_{2} \right) &=& 9\omega_{1}^{2} -
14\omega_{1}\omega_{2}+9\omega_{2}^{2}. \label{5.049b}
\end{eqnarray}
\end{subequations}
\end{widetext}
For the case considered ($\omega_{1}$ and $\omega_{2}$ positive) the solutions
$K_{s}^{\pm,+}$ are always complex conjugate pure imaginary numbers, and do not
represent
a physical solution. On the other hand the solutions $K_{s}^{\pm,-}$ are always
real, but, since $K_{s}^{-,-}$ is always negative the critical coupling at
synchronization is given by
\begin{equation}
K_{s} = K_{s}^{+,-} \left(\omega_{1},\omega_{2} \right). \label{5.050}
\end{equation}
In the figures we have taken $\omega_{1}=1$ and $\omega_{2}=1/3$ for which
$K_{s}\approx 0.734$.

With this two examples of small systems ($N=3$ and $N=4$) we gave a description
of how analytical expressions for the location of the fixed points and the
critical synchronization coupling may be obtained, which we summarize here:
starting from equation (\ref{5.018}), if we make the trigonometric expansion of
the $\sin{\phi_{N}}$ term into sines and cosines of all the phase differences,
then by squaring the two sides of the equation (with subtle rearrangements of
the terms) it is possible to use equations (\ref{5.017}) to write a polynomial
equation either for $z=\sin{\phi_{n^{\ast}}}$ or $x=K^{-1} $. Ultimately the
fixed points will come as the roots of the polynomial, and the critical
synchronization coupling $K_{s}$ appears as the first minimum of the function $K
\left(\phi_{n^{\ast}} \right)$, computed with respect to either
$\phi_{n^{\ast}}$ or $\sin{\phi_{n^{\ast}}}$, as long as $\phi_{n^{\ast}}$ is
chosen properly (for further information about this point see \cite{tilles}).
The problem with this approach is that the degree of the polynomial increases
with the number of oscillators, and for $N \geq 5$ the degree is typically too
high to obtain analytical roots. Nevertheless if we make use of highly
symmetrical configurations of natural frequencies the behavior of the system may
be analytically explored to a large extent. This is the task that we shall
undertake in the next section.

\section{Two natural frequencies}

The simplest system with nontrivial behavior one may consider is composed of
only
two natural frequencies. But as local coupled systems present a strong
dependence on the local configurations, we will get rid of the inhomogeneities
by assuming an organized distribution of the natural frequencies:
\begin{eqnarray}
\omega_{n}= \left\{
\begin{array}{ll}
\omega, & \textrm{for $n$ odd} \\
-\omega, & \textrm{for $n$ even}
\end{array} \right. \label{5.051}
\end{eqnarray}
In the synchronized region the symmetry reduces the equations to a set of
constraints on the phase differences $\phi_{n}$,
\begin{eqnarray}
\sin{\phi_{n}}= \left\{
\begin{array}{ll}
\sin{\phi_{1}}, & \textrm{for $n=3,5,...,N-1$,} \\
\sin{\phi_{2}}, & \textrm{for $n=4,6,...,N-2$.}
\end{array} \right., \label{5.052}
\end{eqnarray}
while $\phi_{1}$ and $\phi_{2}$ satisfy
\begin{subequations}
\label{5.053}
\begin{eqnarray}
\hspace{-0.7cm}
\sin{\left[\sum_{n=\textrm{odd}}^{N-1}\phi_{n}\left(\phi_{1}\right)
+ \hspace{-0.2cm} \sum_{n=\textrm{even}}^{N-2}\phi_{n}\left(\phi_{2}\right)
\right]}
+\sin{\phi_{1}} &=& \frac{\omega}{K}, \label{5.053a} \\
\hspace{-0.7cm} \sin{\phi_{1}} - \sin{\phi_{2}} &=& \frac{\omega}{K}.
\label{5.053b}
\end{eqnarray}
\end{subequations}
A straightforward solution of this set corresponds to take $\phi_{n}=\phi_{1}$
for $n$ even, $\phi_{n}=\phi_{2}$ for $n$ odd and $\phi_{2}=-\phi_{1}$,  such
that (\ref{5.053}) can be written as
\begin{equation}
\sin{\phi_{1}} = \frac{\omega}{2K}. \label{5.054}
\end{equation}
As this set of solutions ($\phi_{n}$) maximizes the left hand side of
(\ref{5.053a}), the critical synchronization coupling is determined as the first
solution of equation (\ref{5.054}), i.e.,
\begin{equation}
K_{s} = \frac{\omega}{2}. \label{5.055}
\end{equation}
Although this specific choice of the phases gives a fixed point solution valid
for $K \geq K_{s}$, the synchronized region is full with solutions obtained from
combinations of $\phi_{n}^{+}=\phi_{1,2}$ and $\phi_{n}^{-}=\pi-\phi_{1,2}$, in
a way that a complete description of the system involves all possible
combinations of the solutions to equation (\ref{5.052}).

To obtain all the solutions (fixed points) with a prescribed
symmetry, we will explore the system's
symmetries to get the simplest description of the dynamics.
Since equation (\ref{5.052}) tells us that there are two independent phases, it
is possible to consider proper initial conditions $\phi_{n} \left(t=0\right)$
that reduce the original $\left(N-1\right)$-dimensional system down to a
$2$-dimensional set of equations. If we define
\begin{subequations}
\label{5.056}
\begin{eqnarray}
\phi_{n}^{+} &=& \left\{
\begin{array}{ll}
\phi_{1}, & \textrm{for $n=3,5,...,N-1$,} \\
\phi_{2}, & \textrm{for $n=4,6,...,N-2$.}
\end{array} \right., \label{5.056a} \\
\phi_{n}^{-} &=& \left\{
\begin{array}{ll}
\pi - \phi_{1}, & \textrm{for $n=3,5,...,N-1$,} \\
\pi - \phi_{2}, & \textrm{for $n=4,6,...,N-2$.}
\end{array} \right., \label{5.056b}
\end{eqnarray}
\end{subequations}
the equations of motion are written as
\begin{subequations}
\label{5.057}
\begin{eqnarray}
\dot{\phi_{1}} &=& 2 \omega -K \left(2 \sin{\phi_{1}} - \sin{\phi_{2}}
+\sin{\Phi}\right), \label{5.057a} \\
\dot{\phi_{2}} &=&-2 \omega + 2K \left(\sin{\phi_{1}} - \sin{\phi_{2}} \right),
\label{5.057b}
\end{eqnarray}
\end{subequations}
where
\begin{eqnarray}
\hspace{-0.5cm} \Phi &\equiv& \Phi^{\pm,\pm,... }
\left(\phi_{1},\phi_{2}\right)= -\phi_{N} \nonumber \\
\hspace{-0.5cm} &=& \phi_{1} + \phi_{2} +\sum_{n=3,5,...}^{N-1}\phi_{n}^{\pm}
\left(\phi_{1}\right) +\sum_{n=4,6,...}^{N-2}\phi_{n}^{\pm}
\left(\phi_{2}\right),
\end{eqnarray}
defines each of the $2^{N-3}$ solutions that compose the $\phi_{N}$ term, thus
generating all the subspaces of the system. The stability of the solutions is
described by the eigenvalues of the Jacobian matrix
\begin{equation}
\lambda_{\pm} = \frac{\tau \pm \sqrt{\tau^{2} - 4\Delta} }{2}, \label{5.059}
\end{equation}
where $\tau$, the trace and $\Delta$, the determinant are
defined by
\begin{subequations}
\label{5.060}
\begin{eqnarray}
\tau &=& - K \left[2\left(\cos{\phi_{1}}+\cos{\phi_{2}}\right) + \cos{\Phi}
\partial_{\phi_{1}}\Phi  \right], \label{5.060a} \\
\Delta &=& 2 K^{2} \left[ \cos{\Phi \left(\cos{\phi_{2}} \partial_{\phi_{1}}\Phi
- \cos{\phi_{1}} \partial_{\phi_{2}}\Phi \right) } \right. \nonumber \\
&& \left. + \cos{\phi_{1}}\cos{\phi_{2}} \right].  \label{5.060b}
\end{eqnarray}
\end{subequations}

\begin{figure}[!t]
\centering
\includegraphics[width=1.0\linewidth]{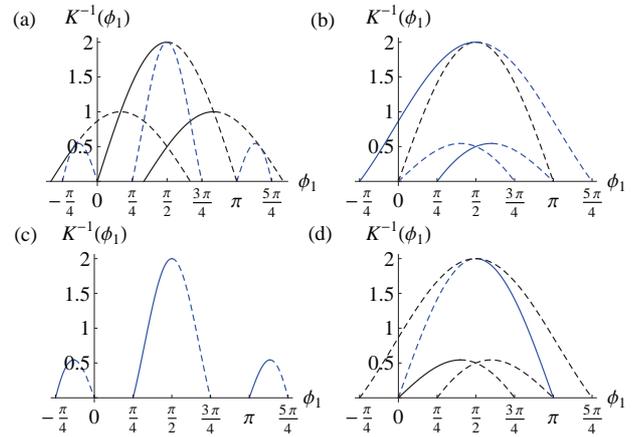}
\caption{(Color online) Bifurcation diagram $K^{-1}\left(\phi_{1}\right)$
showing the solutions and their stability defined in the symmetric spaces for a system with $N=6$  and $\omega=1$. Solid
(dashed) lines represent stable (unstable) solutions. (a) Subspace $\Phi^{+++}$.
(b) Subspace $\Phi^{-++}$. (c) Solutions $\phi_{2}^{-}$ of subspace
$\Phi^{+-+}$. (d) Subspace $\Phi^{-+-}$.} \label{Cap05_fig07}
\end{figure}

Starting with a $N=6$ system, the symmetrical subspaces are defined by
$\Phi^{\pm\pm\pm}$ such that the equation (\ref{5.053a}) has the following
representations in each subspace:
\begin{subequations}
\label{5.061}
\begin{eqnarray}
\hspace{-0.8cm} \Phi^{+++} \quad &\rightarrow& \quad
\sin{\left(3\phi_{1}+2\phi_{2}\right)}
+\sin{\phi_{1}}=\frac{\omega}{K}, \label{5.061a} \\
\left.\begin{array}{ll}
\hspace{-0.8cm} \Phi^{-++}  \\
\hspace{-0.8cm} \Phi^{++-}
\end{array} \right\} \quad &\rightarrow& \quad
-\sin{\left(\phi_{1}+2\phi_{2}\right)} +\sin{\phi_{1}}=\frac{\omega}{K},
\label{5.061b} \\
\hspace{-0.8cm} \Phi^{+-+} \quad
&\rightarrow& \quad -\sin{3\phi_{1}} +\sin{\phi_{1}}=\frac{\omega}{K},
\label{5.061c} \\
\hspace{-0.8cm} \Phi^{-+-} \quad
&\rightarrow& \quad \sin{\left(-\phi_{1}+2\phi_{2}\right)}
+\sin{\phi_{1}}=\frac{\omega}{K},  \label{5.061d} \\
\left.
\begin{array}{ll}
\hspace{-0.8cm} \Phi^{--+} \\
\hspace{-0.8cm} \Phi^{+--} \\
\hspace{-0.8cm} \Phi^{---}
\end{array} \right\}
\quad &\rightarrow& \quad \sin{\phi_{1}} = \frac{\omega}{2K}.  \label{5.061e}
\end{eqnarray}
\end{subequations}
Since the set (\ref{5.061}) corresponds to all possible solutions of equation
(\ref{5.053a}),  every fixed point  in the synchronized region can be
represented by the functions $K^{-1}\left(\phi_{1}\right)$, obtained from
(\ref{5.061}) and (\ref{5.053b}). The solutions, with their corresponding
bifurcations point in each subspace represented by the pair
$\left(K^{\ast},\phi_{1}^{\ast}\right)$, are described in the table
\ref{Table01} and illustrated in figure \ref{Cap05_fig07} with $\omega=1$.
\begin{table}[!t]
\begin{center}
  \begin{tabular}{| c | c | }
    \hline
    \multicolumn{2}{| c |}{Subspace $\Phi^{+++}$} \\
    \hline
    $\phi_{2}^{+} = \frac{2\pi}{3}n_{1}-\phi_{1}$ &
    $\begin{array}{ll}
     n_{1}=0 & \rightarrow  \left(K^{\ast},\phi_{1}^{\ast}\right) =
\left(\omega/2,\pi/2\right)  \\
     n_{1}=1 & \rightarrow  \left(K^{\ast},\phi_{1}^{\ast}\right) =
\left(\omega,5\pi/6\right)  \\
     n_{1}=-1 & \rightarrow  \left(K^{\ast},\phi_{1}^{\ast}\right) =
\left(\omega,\pi/6\right)
    \end{array}$ \\ \hline
    $\begin{array}{c} \phi_{2}^{-} = \pi\left(2n_{2} -1\right) \\ -3\phi_{1}
\end{array}$ & $n_{2}=0 \rightarrow  \left\{ \begin{array}{ll}
    \left(K^{\ast},\phi_{1}^{\ast}\right) = \left(\omega/2,\pi/2\right) \\
    \left(K^{\ast},\phi_{1}^{\ast}\right) \approx \left( 1.84,3.56\right) \\
    \left(K^{\ast},\phi_{1}^{\ast}\right) \approx \left( 1.84,-0.42\right)
    \end{array} \right.$ \\ \hline
    \multicolumn{2}{| c |}{Subspaces $\Phi^{-++}$ and $\Phi^{++-}$} \\
    \hline
    $\phi_{2}^{+} = 2\pi n_{1}-\phi_{1}$ & $n_{1}=0 \rightarrow
\left(K^{\ast},\phi_{1}^{\ast}\right) = \left(\omega/2,\pi/2\right)$ \\ \hline
    $\phi_{2}^{-} = \frac{\pi\left(2n_{2} +1\right)-\phi_{1}}{3}$ &
$\begin{array}{ll}
    n_{2}=0 & \rightarrow \left(K^{\ast},\phi_{1}^{\ast}\right) \approx
\left(1.84,1.88\right) \\
    n_{2}=1 & \rightarrow  \left(K^{\ast},\phi_{1}^{\ast}\right) \approx
\left(1.84,1.26\right) \\
    n_{2}=2 & \rightarrow  \left(K^{\ast},\phi_{1}^{\ast}\right) =
\left(\omega/2,\pi/2\right)
    \end{array}$ \\ \hline
    \multicolumn{2}{| c |}{Subspace $\Phi^{+-+}$} \\
    \hline
    $\begin{array}{ll}
    \phi_{2}^{+} = 3 \phi_{1} \\
    \phi_{2}^{-} = \pi - 3 \phi_{1}
    \end{array}$ & $\begin{array}{ll}
    \left(K^{\ast},\phi_{1}^{\ast}\right) = \left(\omega/2,\pi/2\right) \\
    \left(K^{\ast},\phi_{1}^{\ast}\right) \approx \left(1.84,-0.42\right) \\
    \left(K^{\ast},\phi_{1}^{\ast}\right) \approx \left(1.84,3.56\right)
    \end{array}$ \\ \hline
    \multicolumn{2}{| c |}{Subspace $\Phi^{-+-}$} \\
    \hline
    $\phi_{2}^{+} = \frac{2\pi n_{1}+\phi_{1}}{3}$ & $\begin{array}{ll}
    n_{1}=0 & \rightarrow \left(K^{\ast},\phi_{1}^{\ast}\right) \approx
\left(1.84,1.26\right) \\
    n_{1}=1 & \rightarrow \left(K^{\ast},\phi_{1}^{\ast}\right) \approx
\left(1.84,1.88\right) \\
    n_{1}=-1 & \rightarrow \left(K^{\ast},\phi_{1}^{\ast}\right) =
\left(\omega/2,\pi/2\right)
    \end{array}$ \\ \hline
    $\begin{array}{c} \phi_{2}^{-} = \pi \left(2 n_{2}-1 \right) \\ +\phi_{1}
\end{array}$ & $n_{2}=0 \rightarrow \left(K^{\ast},\phi_{1}^{\ast}\right) =
\left(\omega/2,\pi/2\right)$ \\ \hline
    \multicolumn{2}{| c |}{Subspaces $\Phi^{--+}$, $\Phi^{+--}$ and
$\Phi^{---}$} \\
    \hline
    $ \begin{array}{ll}
    \phi_{2}^{+} = - \phi_{1} \\
    \phi_{2}^{-} = -\pi +\phi_{1}
    \end{array}$ & $\left(K^{\ast},\phi_{1}^{\ast}\right) =
\left(\omega/2,\pi/2\right)$ \\ \hline
  \end{tabular}
  \caption{Solutions for each symmetric subspace and the respective bifurcation
points. The solutions $\phi_{2}^{\pm} \left( \phi_{1}\right)$ for each subspace
(on the left) depend on integer numbers $n_{1,2}$. When the two solutions have
the same bifurcation points $\left(K^{\ast},\phi_{1}^{\ast}\right)$, the
subspace is completely symmetric with respect to $\phi_{2}^{\pm}$ (except for
the stability), and the number of bifurcations is doubled.}
  \label{Table01}
\end{center}
\end{table}

An unexpected feature that comes from the symmetry presented by the system
concerns the characterization of the critical coupling: our analysis shows that
$K_{s}$ is highly degenerate, since each subspace has two bifurcations
responsible for the synchronization, a node-node bifurcation (with opposite
stabilities) coming from the $\phi_{2}^{+}$ solutions, always accompanied by a
saddle-saddle or a saddle-(unstable)node coming from the
$\phi_{2}^{-}$ solutions.  It is important to notice that not
only at $K_{s}$ but at all values of the coupling constant where new solutions
are born there exists the same composition of double bifurcations, as can be
observed in figure \ref{Cap05_fig07}. The reason why this feature comes as
unexpected lies on the behavior of the system when no symmetries are present,
since the critical coupling was always obtained as a single bifurcation and the
presence of double bifurcations was not addressed.

Although the solutions found on the subspaces correspond to all the fixed points
of the system, the stability obtained from the eigenvalues of the bidimensional
Jacobian matrix in equation (\ref{5.059}) does not necessarily correspond to the
stability of the solutions in the complete phase space. Thus,
in order to know the complete stability of solutions it will be necessary to
analyze the eigenvalues from the $\left(N-1\right) \times \left(N-1\right)$
Jacobian obtained from all the equations of motion instead of the bidimensional
set. After carrying out the analysis on the fixed points born at $K_{s}$ we
observed
the existence of only one stable fixed point, originated from $\Phi^{+++}$: the
stable solution from $\phi_{2}^{+}\left(n_{1}=0\right)$. All other fixed points
born at $K_{s}$ were created from bifurcations saddle-saddle like in the
complete phase space.

\begin{figure}[!t]
\centering
\includegraphics[width=1.0\linewidth]{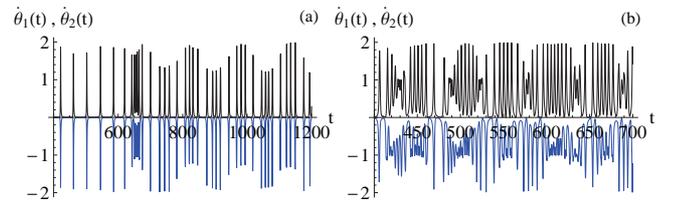}
\caption{Time evolution of the instantaneous frequencies
$\dot{\theta}_{1}\left(t\right)$ and $\dot{\theta}_{2}\left(t\right)$ just below
synchronization ($K=0.499$) for a system with $N=6$ and specular
symmetry with two
natural frequencies and $\omega=1$. (a) For a set of initial
conditions chaotic behavior of
 the time difference between slips is observed, with Lyapunov exponent approximately equal to 0.03.
 (b)  Time evolution of the
instantaneous frequencies for a set of oscillators which wander around phase
space and only eventually
  find the ghost of the bifurcation.}
\label{Cap05_fig08}
\end{figure}

The result of the stability analysis computed in the region slightly above
$K_{s}$ may be summarized as follows: from a total of $32$ solutions created
from $16$
different bifurcations (remember that there exist 8 subspaces of symmetric
solutions, each with two bifurcations), there is one stable fixed point
originated from node-node bifurcations while the other $30$ are saddles. If we
consider a system with such a large number of saddles and one stable node, it
would be only natural to expect an unusual behavior. With that in mind, we
analyzed the time
evolution of the instantaneous frequencies which is illustrated in figure
\ref{Cap05_fig08}. When synchronization is about to happen ($K=0.499$) we
expected to observe phase slips, with the period between slips proportional to
$\left(K_{s}-K\right)^{\alpha}$ {\cite{strog01,hassan01}}. Surprisingly what we
observed here is a strong dependence on the initial conditions: from some
initial conditions the trajectories of the oscillators pass
near the regions where the bifurcations appear, but as there are several
possible bifurcations they get shifting from one to another (each producing a
slip of a different strength) and the period between phase
slips behaves chaotically (figure \ref{Cap05_fig08}a). To characterize the behavior of this time difference between phase
slips we calculate the largest Lyapunov exponent
$\lambda$ \cite{rosenstein, abarbanel}  which was found to be approximately
$ 0.03$; at the
same time there are initial conditions that make the oscillators wander around
phase space and only eventually find a neighborhood of a bifurcation, what does
not necessarily happen simultaneously for all oscillators, resulting on erratic
behavior and intermittent phase slips as seen in figure \ref{Cap05_fig08}b.
This unusual behavior is also observed for values of $K$ slightly
above $K_{s}$ not always finding the stable fixed point in a reasonable time.

So far we have described the behavior of the system prior to and on a
neighborhood immediately after the onset of synchronization. Now we will extend
the discussion to the solutions in the synchronized region, that present some
peculiarities coming from the high density of solutions. We shall focus our
attention on the $\Phi^{++...}$ subspace, which present most (if not all) of the
system's features and has a direct link to the system's stability in the whole
phase space which will be explained in the next section.

\begin{figure}[!t]
\centering
\includegraphics[width=1.0\linewidth]{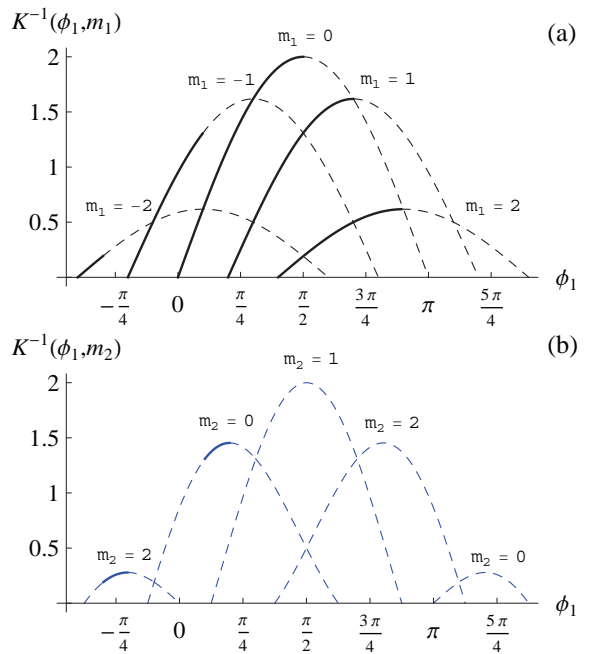}
\caption{(Color online) The plots of $ K_{\pm}^{-1}\left(\phi_1,m_{1,2}\right)$
representing the
fixed points of the system synchronized region with two natural
frequencies and $N=10$. Solid lines
(dashed) represent stable fixed points (unstable). (a) Solutions
of $K_ {+}^{- 1}\left(\phi_{1}, m_ {1}\right)$ with  bifurcations type node-node
$ m_ {1} = 0 $;   saddle-(stable) node bifurcation for $ m_ {1} = 1 $ and $ m_
{1} = $ 2, and saddle- (unstable) node  for
$ m_ {1} =- 1 $ and $ m_ {1} =- 2 $. (b)  solutions
$ K_{-}^{- 1}\left(\phi_{1}, m_{2}\right)$, with saddle-saddle bifurcations
$ m_ {2} = 1 $, saddle-(stable)node  bifurcation in the left of
$ m_ {2} = 0 $ and $ m_ {2} = 2$ ; saddle-(unstable)node  bifurcation in the
right of
$ m_ {2} = 0 $ and $ m_ {2} = 2$ .}
\label{Cap05_fig09}
\end{figure}

For an arbitrary number $N$ of oscillators, the $\Phi^{++...}$ representation of
equation (\ref{5.053a}) is given by
\begin{equation}
\sin{\left[\frac{N \phi_{1} + \left( N-2\right) \phi_{2}}{2} \right]}
+\sin{\phi_{1}} = \frac{\omega}{K}. \label{5.067}
\end{equation}
Combining (\ref{5.067}) with (\ref{5.053b}), the variables $\phi_{1}$ and
$\phi_{2}$ are related through the equation
\begin{equation}
\sin{\left[\frac{N \phi_{1} + \left( N-2\right) \phi_{2}}{2} \right]} =
-\sin{\phi_{2}}, \label{5.068}
\end{equation}
which admits two types of solutions:
\begin{subequations}
\label{5.069}
\begin{eqnarray}
\hspace{-0.5cm} \phi^{+}_{2} &=& \frac{4 \pi m_{1}}{N} - \phi_{1},\qquad
m_{1}=0,\pm1,\pm2,..., \\ \label{5.069a}
\hspace{-0.5cm} \phi^{-}_{2} &=& \frac{2 \pi \left(2 m_{2} -1\right) - N
\phi_{1}}{N-4},\quad m_{2}=0,1,2,... .\label{5.069b}
\end{eqnarray}
\end{subequations}
For each type of solution (\ref{5.069}) there exists a function
$K_{\pm}^{-1}\left(\phi_{1}\right)$ which describes the fixed points of the
system,
\begin{subequations}
\label{5.070}
\begin{eqnarray}
&& \hspace{-1.2cm} K_{+}^{-1}\left(\phi_{1},m_{1}\right) = \nonumber \\
&& \hspace{-1.0cm}  \frac{1}{\omega} \left[\sin{\phi_{1}} \left(1 +
\cos{\frac{4\pi m_{1}}{N}} \right) - \cos{\phi_{1}} \sin{\frac{4\pi m_{1} }{N}}
\right], \label{5.070a} \\
&& \hspace{-1.2cm}  K_{-}^{-1}\left(\phi_{1},m_{2}\right) = \nonumber \\
&& \hspace{-1.0cm}  \frac{1}{\omega} \left\{\sin{\left[\frac{\left(N-2 \right)
\left(2 m_{2} - 1 \right) \pi - N \phi_{1}}{N-4} \right]} + \sin{\phi_{1}}
\right\}, \label{5.070b}
\end{eqnarray}
\end{subequations}
where the minima $\partial_{\phi_{1}}
K_{\pm}\left(\phi_{1}\right)\vert_{\phi_{1}=\phi_{1}^{\ast}}=0$ (maxima of
$K_{\pm}^{-1}$) give the points $\phi_{m_{1}}^{\ast}$ and $\phi_{m_{2}}^{\ast}$
where the bifurcations occur,
\begin{subequations}
\label{5.071}
\begin{eqnarray}
\tan{\phi^{*}_{m_{1}}} &=& - \frac{1 + \cos{\frac{4\pi
m_{1}}{N}}}{\sin{\frac{4\pi m_{1}}{N}} },  \label{5.071a} \\
\cos{\phi^{*}_{m_{2}}} &=&  \frac{N \cos{\left[ \frac{\left(N-2 \right)
\left(2 m_{2} - 1 \right) \pi -N  \phi^{*}_{m_{2}}}{N-4} \right]}}{N-4}.
\label{5.071b}
\end{eqnarray}
\end{subequations}
Figures \ref{Cap05_fig09}a and \ref{Cap05_fig09}b show the solutions
$K_{+}^{-1}\left(\phi_{1},m_{1}\right)$ and
$K_{-}^{-1}\left(\phi_{1},m_{2}\right)$, respectively, with the stable branches
in solid lines for $N=10$. The stability of the
solutions is addressed via the eigenvalues of the 2-dimensional Jacobian matrix,
and we see that the systems exhibits: saddle-(stable)node bifurcations for
$m_{1}=1$ and $2$, (stable)node-(unstable)node for $m_{1}=0$,
saddle-(unstable)node for $m_{1}=-1$ and $-2$ for the case
$K_{+}^{-1}\left(\phi_{1},m_{1}\right)$. For the figure \ref{Cap05_fig09}b we
obtain: saddle-(stable)node bifurcations for $m_{1}=0$ and $2$ on the left,
saddle-(unstable)node for $m_{2}=0$ and $2$ on the right, and a saddle-saddle
$m_{2}=1$.

\begin{figure}[!t]
\centering
\includegraphics[width=1.0\linewidth]{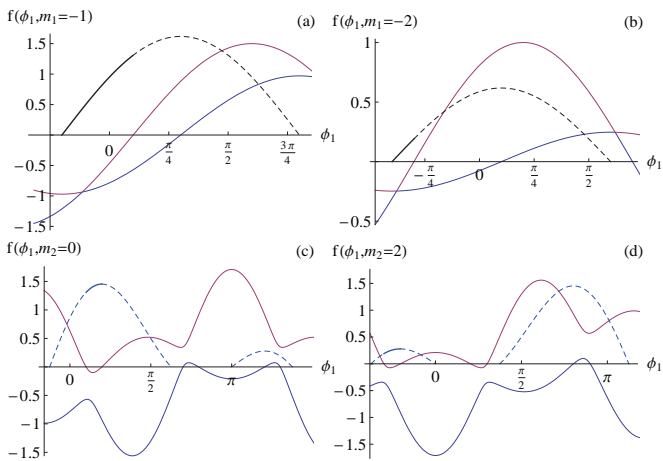}
\caption{(Color online) Branches of solutions for the system with
$N=10$ and two natural
frequencies,  that have changed the
stability by varying the strength of the coupling K. Stable solutions (unstable)
are represented by solid lines (dashed). Upper continuous curves (purple)
represents the real part of $\lambda_{-} \left( \phi_{1} \right)$, and down
curves (blue) represent the real part of $\lambda_{+} \left( \phi_{1} \right)$.
Figures (a) and (b): bifurcations labeled by $ m_ {1} =- 1 $ and $ m_ {1} =- 2
$, respectively, showing that the solutions are born on the right  are saddles
and become stable node when we the real part of $\lambda_ {-}$ becomes negative.
Figures (c) and (d): bifurcations labeled by $ m_ {2} = 0 $ and $ m_ {2} = $ 2,
respectively, showing that the solutions are born on the left are stable nodes
and become saddles when the real part of $\lambda_{-}$ becomes positive.}
 \label{Cap05_fig10}
\end{figure}

The first feature that draws attention in figure \ref{Cap05_fig09} is the
presence of changes in the stability of a given branch when the coupling varies
(see $m_{1}=-1$ and $-2$ and the left branches for $m_{2}=0$ and $2$). From the
real part of the eigenvalues (figure \ref{Cap05_fig10}) we could see that
$\lambda_{-}\left( \phi_{1} \right)$ is responsible for the inversions: at the
bifurcations of $m_{1}=-1$ and $-2$ the fixed point is born as a saddle
($\textrm{Re}\left[\lambda_{+}\right]<0$ and
$\textrm{Re}\left[\lambda_{-}\right]>0$), but then it changes into a stable node
at the point where the real part of $\lambda_{-}$ becomes negative (figures
\ref{Cap05_fig10}a and \ref{Cap05_fig10}b). For the bifurcations on the left of
$m_{2}=0$ and $2$, shown in figure \ref{Cap05_fig09}b, the real part of
$\lambda_{-}$ is positive and becomes negative after a given value of
$\phi_{1}$, where the saddles become stable nodes (figures \ref{Cap05_fig10}c
and \ref{Cap05_fig10}d).

\begin{figure}[!t]
\centering
\includegraphics[width=1.0\linewidth]{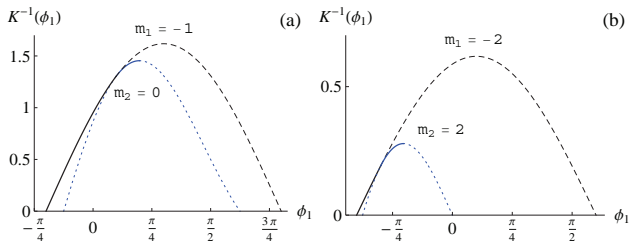}
\caption{(Color online). Region where solutions exchange stability via a
transcritical bifurcation for the cases
$ K_{\pm}^{-1} \left(\phi_{1} \right) $. (a)   Stable node  denoted by $ m_ {2}
= 0 $ loses
stability to the saddle  denoted by $ m_ {1} =- 1 $ at the intersection of the
solutions. (b) Stable node
 $ m_ {2} =2 $ loses stability to the saddle of $ m_ {1} =- 2 $ in
intersection of the solutions.Solid(dashed) lines correspond to stable(unstable)
solutions.(Both figures are done for a system with two natural
frequencies and
$N=10$)} \label{Cap05_fig11}
\end{figure}

The exact value of the points where the branches change stability can be
identified if we analyze all those solutions simultaneously as in figure
\ref{Cap05_fig11}: the branch born from the bifurcation on the left of $m_{2}=0$
loses stability when it touches the branch born at $m_{1}=-1$ for $K \approx
0.764$ with $\phi_{1}=\pi/10$, which becomes stable when the coupling increases
(figure \ref{Cap05_fig11}a); the same process occurs with the branch coming from
the bifurcation on the left of $m_{2}=2$, which exchanges stability with the
$m_{1}=-2$ solution at $K \approx 5.236$ and $\phi_{1}=-3\pi/10$ (figure
\ref{Cap05_fig11}b). The conclusion from this analysis is that a transcritical
bifurcation is indeed responsible for the changes in the stability of the fixed
points.

Changes in the stability properties of the fixed points occur in the stable as
well as the unstable branches as can be seen in the figures \ref{Cap05_fig10}c
and \ref{Cap05_fig10}d: at the saddle-(unstable)node bifurcations on the right
of $m_{2}=0$ and $m_{2}=2$, the unstable nodes become saddles when the real part
of $\lambda_{+}$ becomes negative. Since the solutions
$K_{\pm}^{-1}\left(\phi_{1}\right)$ are symmetric with respect to the
$\phi_{1}=\pi/2$ axis, the mechanism which alters the stability of those
solutions is the same process as described before, with the exchanging points
given by $\phi_{1}=9\pi/10$ and $\phi_{1}=13\pi/10$.

The stability exchange via a transcritical bifurcation may also take place on
the complete phase space of the system, although the exchange is not necessarily
made between
solutions of the same subspaces. Therefore to obtain all the stable solutions of
the system, independent of the size of their basins of attraction, it is
necessary to obtain all the points in space where two (or more) solutions
collide and to analyze the stability of those fixed points before and after the
collision (while at the same time the stability of all the other solutions
should be considered). Since the number of solutions is exceedingly large to
handle, even for small systems, this type of symmetry may not be the most
enlightening choice to obtain a description of the synchronized region in a
simple (analytical) fashion. So in the next section we will restrict our
symmetry assumptions to more general cases, looking for the analyticity limits
of the model.

\section{Specular symmetry}

\begin{figure}[!t]
\centering
\includegraphics[width=1.0\linewidth]{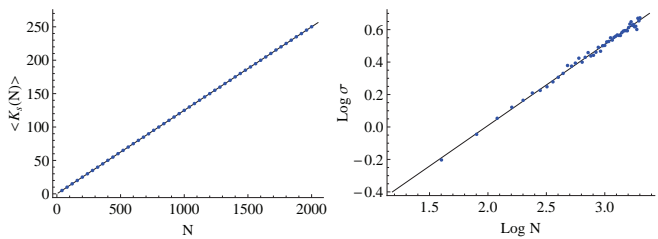}
\caption{Left: behavior of $ \left \langle K_ {s} {\left( N\right)} \right
\rangle $ for natural frequencies generated by a uniform distribution with $ a =
0 $ and $ b =1 $ and systems with specular symmetry. Right:
standard deviation $ \sigma \left[K_{s} \left(N \right)
\right]$ from the numeric outcome in logarithmic scale. The dots correspond to
the result obtained through the simulation,  with average obtained from 1000
instances. The black line correspond to the curve  calculated  via
(\ref{5.077}).} \label{Cap05_fig14}
\end{figure}

Let us consider an even number of oscillators with a configuration of
frequencies that satisfy the following symmetry:
\begin{equation}
\omega_{N/2+n} =-\omega_{n}, \quad n=1,..,N/2,  \label{5.072} \\
\end{equation}
and also consider that all natural frequencies in the interval $\left[\omega
_{1},\omega_{N/2}\right]$ are positive. If we write the variables $\phi_{n}$ in
terms of $\phi_{N/2}$,
\begin{subequations}
\label{5.078}
\begin{equation}
\sin{\phi _{n}} =  \sin \phi _{N/2} - \frac{1}{K} \sum_{j=n+1}^{n^{\ast }}
\omega _{j}, \label{5.078a}
\end{equation}
for $n=1,...,N/2 -1$ and
\begin{equation}
\sin{\phi _{n}} = \sin \phi _{N/2} + \frac{1}{K}\sum_{j=N/2+1}^{n} \omega_{j},
\label{5.078b}
\end{equation}
\end{subequations}
for $n=N/2+1,...,N-1$, the synchronized region is determined by
the solutions of equation (\ref{5.018}) in the form:
\begin{equation}
\sin{\phi_{N/2}}  - \sin{\phi_{N}  \left(
\phi_{N/2}\right)}=\frac{\sum_{n=1}^{N/2}\omega_{n}}{K}. \label{5.075}
\end{equation}
From the analysis of the symmetry properties it is possible to
conclude that there is a solution of this equation with
$\phi_{n+N/2}=-\phi_{n}$ as long as $\sin{\phi_{N}}=-\sin{\phi_{N/2}}$. Since
the last identity maximizes the left hand side of equation (\ref{5.075}), the
critical synchronization coupling is given by
\begin{equation}
K_{s}= \frac{\sum_{j=1}^{N/2} \omega_{j}}{2}. \label{5.073}
\end{equation}
The determination of the critical coupling
with this symmetry is reduced to adding the first $N/2$ values
of the natural frequencies, therefore any prescription that we choose to
determine the natural frequencies $\omega_{n}\left(N\right)$
($n=1,...,N/2$) will allow us to obtain the dependence of $K_{s}$ on the size of
the system $N$. For instance, if we consider a general case
where the frequencies in the interval $\left[ \omega_{1},\omega_{N/2}\right]$
are obtained from a uniform distribution defined on the
interval $\left[a,b\right]$, with $a>0$ and $b>a$, it is easy to compute the
mean of $K_{s}\left(N\right)$ and the standard deviation from equation
(\ref{5.073}):
\begin{equation}
\left\langle {K_{s} \left( N\right)}\right\rangle=\frac{a+b}{8} N, \quad
\sigma\left[K_{s} \left(N \right) \right] = \frac{b-a}{4 \sqrt{6}} \sqrt{N}.
\label{5.077}
\end{equation}
These results are corroborated by the simulation as illustrated in figure
\ref{Cap05_fig14}. A statistical study of the solutions for the
local
Kuramoto model with quenched disorder has not yet been addressed in the
literature.
Here we show the average critical coupled strength $K_s$ when the random field
have specular symmetry. The analytical solutions for this systems depends on the
possibility to identify the maximum sequential sum of the frequencies inside the
ring \cite{tilles}. Analytical solutions for $K_s$ can be obtained for all
symmetries that
allow this identification.

Although the imposition of the symmetry defined in (\ref{5.072}) on the set of
natural frequencies allowed us to determine $K_{s}$ analytically, there is still
the problem of finding all the solutions in the synchronized
region. The selection of the phase difference $\phi_{N/2}$ as the independent
variable warrants that no sum in the equations (\ref{5.078}) will be larger than
the sum on the right hand side of (\ref{5.075}). From this we
infer that there exists a region of values of $K$ such that
the set (\ref{5.078}) has a solution (the right hand side of each equation
always belongs to the interval $\left[-1,1\right]$) while
$\phi_{N/2}$ does not satisfy (\ref{5.075}). This means that all the solutions
in the synchronized region exist after a value $K=K_{s}$ which guarantees the
existence of a solution for equation (\ref{5.075}), while in the synchronized
region ($K\geq K_{s}$) the solutions of the form  $\phi_{N/2}\left(K\right)$
correspond to the fixed points of the system.

To get all the solution in terms of  $\phi_{n}\left(\phi_{N/2} \right)$ we have
to consider the equations (\ref{5.078}) that define relations between the
phases, thus all possible combinations:
\begin{subequations}
\label{5.080}
\begin{eqnarray}
\phi_{n}^{+} &=& \arcsin{\left(\sin \phi _{N/2} - \frac{1}{K}
\sum_{j=n+1}^{n^{\ast }} \omega _{j}\right)}, \label{5.080a} \\
\phi_{n}^{-} &=& \pi - \arcsin{\left(\sin \phi _{N/2} -
\frac{1}{K} \sum_{j=n+1}^{n^{\ast }} \omega _{j}\right)},
\label{5.080b}
\end{eqnarray}
for $n=1,...,N/2 -1$, and
\begin{eqnarray}
\phi_{n}^{+} &=& \arcsin{\left(\sin \phi _{N/2}  +
\frac{1}{K}\sum_{j=N/2+1}^{n} \omega _{j}\right)}, \qquad ,
\label{5.080c} \\
\phi_{n}^{-} &=& \pi - \arcsin{\left(\sin \phi _{N/2}  +
\frac{1}{K}\sum_{j=N/2+1}^{n} \omega _{j}\right)},
\label{5.080d}
\end{eqnarray}
\end{subequations}
for $n=N/2+1,...,N-1$, should be considered a priori
\footnote{A certain amount of care should be considered when
dealing with equation (\ref{5.080}): the solutions $\phi_{n}^{-}$ were written
as $\pi - \arcsin\left(*\right)$ just for a easier
presentation. It is necessary to know the quadrant where the argument is
located to obtain the correct solution in the opposite quadrant that also
satisfy (\ref{5.078}).}. In order to correctly describe the
systems we shall start with simple frequency configurations.

\begin{figure}[!t]
\centering
\includegraphics[width=1.0\linewidth]{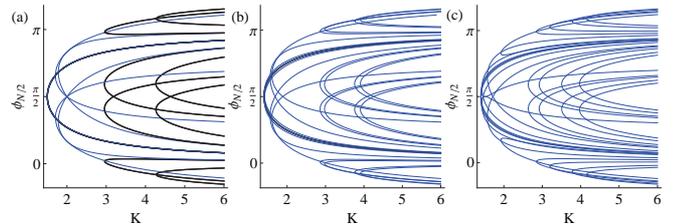}
\caption{(Color online) Numerical solution of equation (\ref{5.080}) in the
space $\phi_{N/2} \times K$ representing the fixed points of the system with $N
=6$ oscillators. (a) Setting
$\omega_{n} =\omega$ for $\left(\omega_{1},
\omega_{2}, \omega_{3}\right)$ with high degeneracy and specular symmetry. (b) The presence symmetry
given by (\ref{5.072}) separates the solutions such that only the bifurcations
located in $K_{s}$ are degenerated. The natural frequencies are $ \omega_{1} = $
1, $ \omega_{2} = 0.7 $ and$ \omega_{3} = 1.1 $.(c) The continuous symmetry
breaking process separates the spectrum of solutions completely, destroying the
degeneracy. The natural frequencies are  $ \omega_{1} = 1 $, $\omega_{2} = 0.7
$, $ \omega_{3} = 1.1 $,  $ \omega_{4} =- 0.8 $, $ \omega_{5} =- $ 0.75 and
$\omega_{6} =- $ 1.25.} \label{Cap05_fig15}
\end{figure}

The simplest configuration of frequencies satisfying the symmetry (\ref{5.072})
is $\omega_{n}=\omega$ for $n=1,...,N/2$, with critical coupling given by
\begin{equation}
K_{s}\left(N\right) = \frac{\omega N}{4}. \label{5.081}
\end{equation}
If we consider a small system $N=6$, all $2^{N-2}=16$
combinations of the kind of solutions (\ref{5.080}) may be
considered to solve (\ref{5.075}) numerically, as shown in figure
\ref{Cap05_fig15}a.
Even though it is possible to determine all solutions for
$\phi_{N/2}=\phi_{3}$ above $K_{s}$, the high symmetry level
presents the same problem of degeneracies in the $\phi_{N/2}
\times K$ space found in the previous section: there are several values of phase
locking $\phi_{1}$, $\phi_{2}$, $\phi_{4}$  and $\phi_{5}$ which lead to the
same result for $\phi_{3}$. In larger systems the process of
assigning an observed fixed point to a phase locked solution is practically
impossible. However there is no need for special prescription for the natural
frequencies within $\left[\omega_{1},\omega_{N/2} \right]$ if we are looking for
solutions
that satisfy (\ref{5.072}).
Even if we consider a case with complete broken symmetry,
equation (\ref{5.075}) will still be valid as long as the sum in the right hand
side continues to be larger than all other sums in
(\ref{5.078}).

Since small perturbations in the natural frequencies have little effect in the
synchronized region, it is possible to start from the configuration that
produces equation (\ref{5.081}) and continuously vary the
frequencies while following the solutions in the $\phi_{N/2}
\times K$ space. If we do this while preserving the specular symmetry
(\ref{5.072}), we observe that all solutions born
after $K_{s}$ are separated (as can be seen in figure \ref{Cap05_fig15}b for
$\omega_{1}=1$, $\omega_{2}=0.7$ and $\omega_{3}=1.1$). If we continue this
process until the configurations satisfy only equation
(\ref{5.075}) (without symmetries), the degeneracy of the solutions is destroyed
(the figure \ref{Cap05_fig15}c shows a case with $\omega_{1}=1.0$,
$\omega_{2}=0.7$, $\omega_{3}=1.1$, $\omega_{4}=-0.8$, $\omega_{5}=-0.75$ and
$\omega_{6}=-1.25$, where several solutions exist above
$K_{s}$ but without any overlaps).

The continuous symmetry breaking of equation (\ref{5.072})
illustrates a property present in any configuration of natural frequencies: the
presence of degeneracies in the space $\phi_{n^{\ast}} \times K$
is a consequence of the symmetries in the set
$\left\{\omega\right\}_{N}$. The importance of this property
becomes evident when one consider a system with asymmetric frequency
configuration: in general the main interest does not lie on the description of
the full set of solutions above $K_{s}$, but on the determination of the stable
fixed points. While for a chain of oscillators the stable fixed point is
characterized by $\cos{\phi_{n}} >0$ (for all $n$), the loop structure of the
ring allows the existence of stable fixed points with $\cos{\phi_{n}} <0$ for
some $n$ (see Lee \emph{et al.} \cite{lee}), and our recent work \cite{tilles}
showed that these phase differences may either be $\phi_{n^{\ast}}$ or
$\phi_{N}$.
Hence it is only necessary to analyze the solutions of (\ref{5.075}) with
relations $\phi _{n}\left(\phi_{N/2}\right)$ defined by (\ref{5.080a}) and
(\ref{5.080c}), i.e., $\phi _{n} \in \left[-\pi,\pi\right]$ for
$n \neq n^{\ast}$, while all other combinations of (\ref{5.080}) always lead to
unstable solutions. Now it becomes clear the choice for the
$\Phi^{++...}$ in the previous chapter: all stable fixed points for the dynamics
on the whole phase space come from the solutions of this specific subspace.

A general characterization of the synchronized region in the
absence of symmetries (or with a small amount) was already done in our previous
work \cite{tilles}, therefore it is not necessary to do it again. Nevertheless
we can use the symmetry properties of the natural frequencies in order to study
the behavior of the critical synchronization coupling $K_{s}$ as a function of
the number of oscillators, as discussed in the next section.

\section{Asymptotic behavior of the critical coupling }

Possibly the simplest non-degenerate case analyzable through
the solutions of (\ref{5.075}) consists in the configuration where the
frequencies are evenly spaced in the interval $[-\gamma
,\gamma ]$:
\begin{equation}
\omega _{n}=\frac{\gamma }{N-1}\left( -2n+N+1\right), \qquad n=1,...,N.
\label{5.082}
\end{equation}
The frequencies of the oscillators are mirror images with respect to the axis
between $\phi_{N}$ and $\phi_{N/2}$, i. e., $\omega_{N+1-n} = -\omega_{n}$ for
$n=1,...,N/2$, which gives
\begin{equation}
\sin{\phi_{N/2+n}} = \sin{\phi_{n}}, \qquad n=1,...,N/2-1. \label{5.083}
\end{equation}
Assuming $\phi_{N/2+n} = \phi_{n}$ the phase differences $\phi_{n}$ in
(\ref{5.080}) are given by:
\begin{equation}
\phi _{n}\left( \phi _{N/2},K\right) = \arcsin{\left[ \sin
\phi_{N/2}-\frac{\gamma \left(N-2 n\right)^{2}}{4\left( N-1\right) K} \right]},
\label{5.084}
\end{equation}
for $n=1,...,N/2-1,$ such that the solutions in the synchronized region are
obtained from
\begin{equation}
\sin \left( \phi _{N/2}+2\sum_{n=1}^{\frac{N}{2}-1}\phi _{n}\right) +\sin \phi
_{N/2}=\frac{\gamma N^{2}}{4\left( N-1\right) K}. \label{5.085}
\end{equation}

\begin{figure}[!t]
\centering
\includegraphics[width=1.0\linewidth]{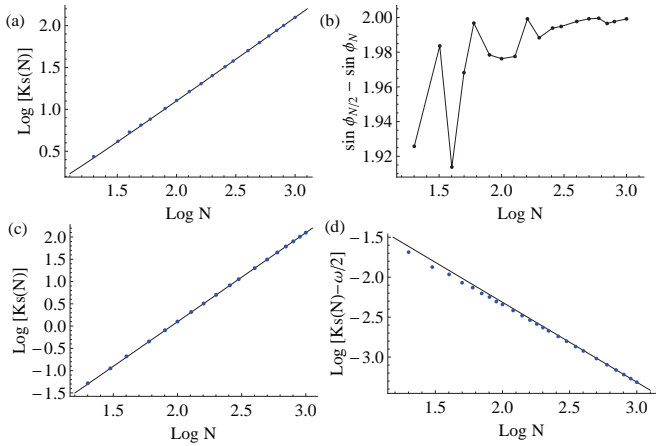}
\caption{Behavior of $K_{s} \left(N \right)$ obtained from different
configurations of natural frequencies, represented by dots, and the continuous
line are the theoretical asymptotic curves: a) $\omega_{n}$ is defined in
(\ref{5.082}) with $\gamma =1$ (logarithmic scale) and the line is given by
equation (\ref{5.087}). b) Behavior of $\sin{\phi_{N/2}} - \sin{\phi_{N}}
\approx 2$. c) $\omega_{n}$ is defined in (\ref{5.088}) with $\epsilon = 0.002$
(logarithmic scale) and the line is given by the equation (\ref{5.091}). d)
$\omega_{n}$ is defined in (\ref{5.093}) with $\gamma = 1$, and the coupling is
inversely proportional to the system size. The curve corresponds to the
asymptotic fitting (\ref{5.096}) with $ A = $ 0,485.}
\label{Cap05_fig17}
\end{figure}

The bifurcation responsible for the full synchronization is always the first
solution of (\ref{5.085}) and it can appear near any of the two
boundaries of the solvability region, i.e.,
it is characterized either by $\sin{\phi_{N/2}} \approx 1$ or
$\sin{\phi_{N}} \approx -1$. The number of bifurcations giving
birth to stable solutions is proportional to the number of oscillators, and for
increasing $N$ we observed that they tend to pile up near the
minimum value of $K$ under which equation (\ref{5.085}) may
present a solution. In other words, for large $N$ it is possible to approximate
$\sin{\phi_{N/2}} - \sin{\phi_{N}} \approx 2$ to obtain an asymptotic behavior
for the critical coupling:
\begin{equation}
K_{s}^{\textrm{asymptotic}}\left(N\right) = \frac{\gamma N^{2}}{8\left(
N-1\right)}, \label{5.087}
\end{equation}
Figure \ref{Cap05_fig17}a shows the results obtained from the
simulation in contrast to the analytical expression in (\ref{5.087}), while
figure \ref{Cap05_fig17}b exhibits the actual value of the sum of the two sines,
thus corroborating the used approximation.

That asymptotic behavior of $K_{s} \sim N$, also observed in the cases treated
previously, is not just a consequence of the evenly spaced frequencies but
rather from the fact that the interval is finite, thus the differences
$\omega_{n}-\omega_{n+1}$ are proportional to $1/N$. If we consider the
configuration
\begin{equation}
\omega _{n}=\frac{\epsilon }{2}\left( -2n+N+1\right), \qquad n=1,...,N,
\label{5.088}
\end{equation}
where $\omega_{n}-\omega_{n+1} = \epsilon$, independent of $N$, the equations
(\ref{5.084}) and (\ref{5.085}) are written as
\begin{equation}
\phi _{n}\left( \phi _{N/2},K\right) = \arcsin{\left[ \sin
\phi_{N/2}-\frac{\epsilon \left(N-2n\right)^{2}}{8 K}\right]}, \label{5.089}
\end{equation}
for $n=1,...,N/2-1$, and
\begin{equation}
\sin \left( \phi _{N/2}+2\sum_{n=1}^{\frac{N}{2}-1}\phi _{n}\right) +\sin \phi
_{N/2}=\frac{\epsilon N^{2}}{8 K}. \label{5.090}
\end{equation}
Since this system presents the same characteristics as the
latter, the asymptotic limit of $K_{s}\left(N\right)$ can also
be obtained with
 $\sin{\phi_{N/2}} - \sin{\phi_{N}} \approx 2$, so
\begin{equation}
K_{s}^{\textrm{asymptotic}}\left(N\right) = \frac{\epsilon N^{2}}{16},
\label{5.091}
\end{equation}
as shown in figure \ref{Cap05_fig17}c.

For all the analyzed cases, the critical value of the coupling constant at
synchronization grows with $N$, which is common in systems with nearest neighbor
interactions between oscillators, and this includes the chain of oscillators
\cite{strog01}. In both cases given by (\ref{5.072}) as well as the mirror
symmetry ($\omega_{N+1-n}= -\omega_{n}$) present in the evenly spaced
configurations of frequencies (\ref{5.082}) and (\ref{5.088}), any prescription
for the distribution of frequencies that has a well defined behavior of the sum
of the frequencies with $N$, that is,
\begin{equation}
\sum_{n=1}^{N/2}\omega_{n}= f\left(N\right), \label{5.092}
\end{equation}
will have (in general) a behavior of the critical coupling  of the form
$K_{s}\left(N\right) \sim \frac{1}{2} f\left(N\right)$, being exact for
(\ref{5.072}) and asymptotic for the other cases. As in general the sum in
(\ref{5.092}) produces a function which increases with $N$, these configurations
will always have $K_{s} \rightarrow \infty$ when $N \rightarrow \infty$. But,
there exists at least one configuration for which this property is not valid: a
system with mirror symmetry $\omega_{N+1-n}=-\omega_{n}$ and natural frequencies
given by
\begin{equation}
\omega_{n}=\frac{\gamma \left(-1\right)^{n+1}}{N-1}\left( -2n+N+1\right), \quad
n=1,...,N/2. \label{5.093}
\end{equation}
Due to the term $\left(-1\right)^{n+1}$ the largest sequential sum
$\sum_{n}\omega_{n}$ that can be obtained corresponds to the largest element of
the set $\left\{\omega\right\}_{N}$ ($\omega_{1}=\gamma$), such that the
equations that describe this system in the synchronized region are:
\begin{equation}
\phi_{n}\left(\phi_{1}\right) = \arcsin{\left\{\sin{\phi_{1}} -\frac{\gamma
\left[N - 2 + \left(-1\right)^{n}\left(N-2n \right) \right]
}{2\left(N-1\right)K} \right\}}, \label{5.094}
\end{equation}
for $n=2,...,N/2$, and
\begin{equation}
\sin{\left[ 2\phi_{1} + 2\sum_{n=2}^{N/2-1}\phi_{n}\left(\phi_{1}\right) +
\phi_{N/2}\left(\phi_{1}\right) \right]} + \sin{\phi_{1}} = \frac{\gamma}{K}.
\label{5.095}
\end{equation}

When the number of oscillators increases the effect in the solvability region is
the same as that
in the cases of evenly spaced frequencies, therefore in the limit $N\rightarrow
\infty$ we must have $\sin{\phi_{1}}-\sin{\phi_{N}} \sim 2$. But as the right
hand side of equation (\ref{5.095}) does not depend on $N$, the critical
coupling $K_{s}$ should decrease with $N$ (since for small systems $K_{s} >
\gamma/2$). Then, if we take the limit of a very large number of oscillators, we
should obtain
\begin{equation}
K_{s}^{\textrm{asymptotic}}\left(N\right) = \gamma/2 +  \frac{A}{N} +
\textrm{O}\left(N^{-2}\right), \label{5.096}
\end{equation}
as can be seen in figure \ref{Cap05_fig17}d, where $A$ is just a constant. The
importance to obtain a configuration of frequencies which produces a behavior
for the critical coupling  $K_{s} \sim 1/N$ in the LCKM resides in the
comparison with systems that possess larger network connectivity. The critical
coupling in the Kuramoto model (globally connected) decreases with $N$ for
natural frequencies defined in a finite interval \cite{pazo}, while locally
connected model seems to diverge in the limit $N \rightarrow \infty$. At first
sight, one would expect a transition in the asymptotic limit for some critical
network connectivity, but our result indicates that, under a proper organization
of the natural frequencies in the LCKM, the two regimes may be connected in a
continuous way, just as the type of interaction considered in \cite{rogers}. We
believe that this subject deserves further investigation.

\section{Conclusions}

In the present work we gave a comprehensive description of the locally coupled
Kuramoto model under the presence of symmetries and explored to a large extent
the analyticity limits of the system. In the following we summarize the salient
features.

Starting with the analysis of small systems, we gave a full description of the
synchronized region, along with the determination of the critical
synchronization coupling for some cases, and we showed how the fixed points may
be obtained from a general configuration of natural frequencies, independent of
the system size.

To keep our attention on the analytical regime of the model, we considered a
highly symmetric configuration, consisting of two natural frequencies: $\omega$
for odd numbered oscillators and $-\omega$ for even numbered ones. Despite of
its apparent simplicity, many unusual and (or) unexpected results were observed.

We found that the stability of the fixed points depend on the number of
dimensions of the manifold where the dynamics is set to happen: while for
2-dimensional systems (obtained from symmetric initial conditions) a large
number of stable fixed points populate the phase space, the complete system is
characterized by just a small amount of stable fixed points.

Several times we chose initial condition with a given symmetry. This
was done to control the final state.
As an example, consider the case of a system with two natural frequencies. The
conditions given by equation (32)
lead to equation (33). We note that the phase differences depend on two
independent phases
$\phi_1$ and $\phi_2$. All the solutions are constructed from all possible
combinations of
 $\phi_n^{+}$ and $\phi_n^{+}$ in equation (37). So given some specific initial
condition prescribed
 by equation (37), the solutions is already determined by equation (39). If we
start with a set of initial conditions
without the symmetry of the system, we cannot anticipate into which solution of
equation (39)
the dynamics will evolve. Imposing the symmetry on the phases from the
beginning, we
select the attractor
 in the corresponding subspace.

At the onset of synchronization it is well known that one should expect periodic
phase slips, however in the two frequencies domain the system exhibits a strong
dependence on the initial conditions: a regime of phase slips with irregular
period (apparently chaotic) coexists with complete erratic bursting behavior
characterized by a long time absence of phase slips as well as intermittent
appearances. This local character of the system also extends to the synchronized
region, where we could observe that the stable fixed points created slightly
above $K_{s}$ behave only as local attractors in the phase space. The
coexistence of stable fixed points may lead to the appearances of attractor
crowding, as described by Wiesenfeld and Hadley \cite{wiesen}, and we hope
further investigation will clarify the issue.

We also observed stability exchange between synchronized solutions. This feature
happens when two fixed point collide in the phase space. The values of $\phi$
for these exchanges to occur and the nature of each fixed point were determined
through the analysis of the Jacobian's eigenvalues in the reduced system, and
the same feature is expected to happen in the complete phase space.

The presence of a type of specular symmetry enabled us to obtain some analytical
expressions for the critical synchronization coupling, but one important result
came from the analysis of a continuous symmetry breaking of the system: the
degeneracies in the $\phi_{n^{\ast}} \times K$ space are lifted in the absence
of symmetries on the natural frequencies. As a consequence we could show how
just a small amount of the solutions will be stable, and the search for stable
fixed points may be reduced to the analysis of just the solutions composed by
phase differences belonging to the interval $\left[-\pi,\pi\right]$.

By further exploring the symmetry properties we could manage to obtain some
asymptotic results concerning the behavior of the critical synchronization
coupling as a function of the system size $N$. While for general configuration
the LCKM tends to be proportional to the number of oscillators, we found a
counterexample where $K_{s}$ actually decreases with $N$.

Many of the phenomena and results discussed here were not observed in a
completely random natural frequencies distribution. The study of all theses
symmetric cases have brought insights about the complexity of the LCKM and
helped us advance in the understanding and the development of analytical methods
to analyze this kind of system.

\acknowledgments

P.F.C.T. acknowledges support by FAPESP and CAPES (Brazil).

\end{document}